# Security in 5G Networks – How 5G networks help Mitigate Location Tracking Vulnerability

Abshir Ali

Department of Computer Science and Engineering

The Ohio State University

Columbus, OH, United States

ali.752@osu.edu

Guanqun Song

Department of Computer Science and Engineering

The Ohio State University

Columbus, OH, United States

song.2107@osu.edu

Ting Zhu

Department of Computer Science and Engineering

The Ohio State University

Columbus, OH, United States

zhu.3445@osu.edu

*Abstract*— As 5G networks become more mainstream, privacy has come to the forefront of end users. More scrutiny has been shown to previous generation cellular technologies such as 3G and 4G on how they handle sensitive metadata transmitted from an end user mobile device to base stations during registration with a cellular network. These generation cellular networks do not enforce any encryption on this information transmitted during this process, giving malicious actors an easy way to intercept the information. Such an interception can allow an adversary to locate end users with shocking accuracy. This paper investigates this problem in great detail and discusses how a newly introduced approach in 5G networks is helping combat this problem. The paper discusses the implications of this vulnerability and the technical details of the new approach, including the encryption schemes used to secure this sensitive information. Finally, the paper will discuss any limitations to this new approach.

## I. Introduction

Security in 5G networks has become an increasingly important topic as the infrastructure is quickly adopted in the telecommunications industry. It has been projected that by 2028, 5G will account for 4.6 billion subscriptions, more than 50% of all mobile subscriptions worldwide [3]. The evolution of 5G technology is pivotal not only for enhancing network security but also for supporting advanced applications in the Internet of Things (IoT), low-power communication, and autonomous driving. These applications demand robust security measures due to their sensitivity and potential impact on user privacy and safety. For instance, in IoT [18-21], the focus is on securing a multitude of devices with diverse capabilities and communication requirements. Similarly, in low-power communication, ensuring data integrity and security is crucial for devices operating with limited energy resources [22-24]. Additionally, the complexity of concurrent [5-9] and heterogeneous [10-17] communication in 5G networks presents unique security challenges that need to be addressed. While previous generation networks have proven extremely reliable and technologically sound, the lack of security in critical systems have raised questions on whether these flaws can be exploited by malicious actors. One such design flaw in previous generation networks was the concerning lack of encryption in critical metadata being sent from the end user device to the 5G network equipment during initial registration with the network.

The SIM cards in smart phones contain critical information that allow the device to be identified in a 5G network. One of the key pieces of information stored on it is the International Mobile Subscriber Identity (IMSI). This is a 13–15-digit number that allows a device to be uniquely identified on a cellular network [3]. Mobile phones use the IMSI to uniquely identify themselves when they try to connect to a 5G network for the first time. During the initial registration of a device, previous generation networks stipulated that the device must send over the IMSI in plaintext as part of the registration process with the core network [3]. This proved to be a significant security risk since it allowed for third party devices to simply intercept the number before it reached it's intended target. It is during the initial registration with this device that the mobile device will end up sending its unencrypted IMSI.

The dangers of sending an unencrypted IMSI over the network cannot be understated. The most troubling consequence is that it can allow an adversary to track the device across the network [3], given the uniqueness of the number – hence why this vulnerability is known as location tracking vulnerability. It is easy to assume that the technology and cost needed to intercept an unencrypted IMSI would be too high to justify upending both existing and future next generation cellular networks, but it may surprise some that with only hardware costing $7 [3] and a few lines of open-source code, anyone can design a functioning IMSI catcher.

It would be reasonable, therefore, to argue that the solution would be to fully encrypt the IMSI so that it can remain secret during transit. While this is a valid argument, it is not as easy as it may seem. The IMSI is made up of three main components, the Mobile Country Code (MCC), the Mobile Network Code (MNC) and the Mobile Subscriber Identification Number (MSIN) [25]. The MSIN tends to be unique and contain sensitive information and therefore would, without doubt, need to be encrypted The MCC and MNC, however, are used to validate which cellular networks are compatible with the mobile device. As an example, if I have a mobile device with a MCC corresponding with the USA and a MNC corresponding with AT&T, my device would only be able to connect to networks that are in the USA and belong to AT&T. It would not be able to connect to a network that belongs to Verizon. Core networks rely on this information to properly route data. Introducing encryption to these key pieces of information would require extensive redesigning of multiple parts of the network to support decryption of the MCC and the MNC to access this information, which would be costly.

This means that any next generation cellular technology would need to achieve the following:

- It would need to be able to encrypt the IMSI so that the sensitive parts (specifically the MSIN) would be strongly encrypted, while also leaving the MCC and MNC unencrypted to ensure easy routing of data.

- It would need to choose an encryption scheme which will ensure that the partially concealed IMSI cannot be easily decrypted by an adversary.

- It would need to have infrastructure in place which can encrypt the sensitive parts of the IMSI on the sender side and decrypt the partially concealed IMSI on the receiver's side.

This paper will discuss how 5G networks mitigate the risk of location tracking and achieve all three criteria using Subscription Concealed Identifier (SUCI), a modern approach which uses the Elliptical Curve Integrated Encryption Scheme (ECIES), a form of cryptography based on Elliptical Curve Cryptography (ECC), to encrypt the MSIN portion of the IMSI. The paper will give a detailed overview on how the encryption works and how it guarantees the protection of the IMSI from eavesdropping attacks. There will also be a simulation on how the encryption works, both on the sender side and the receiver's side and discuss any known limitations to this protection scheme.

## II. BACKGROUND

To understand how modern 5G networks can help reduce the risk of location tracking vulnerability, we must first discuss the basics on how a cellular device register to a 5G network. As part of the overview, a basic simulation of a 5G network was conducted using the open source 5G core network simulator Open5Gs and UERANSIM, an open-source UE/gNB simulator. The simulations were conducted using two pre-built virtual machines, both of which were obtained from [2].

### A. Key Terminology

a. **User Equipment (UE)** refers to an end user device which connects to a 5G network. This can include devices such as smartphones, laptops, Internet of Things (IOT) devices etc.

b. **Next Generation NodeB (gNodeB/gNB/base station)** refers to the station which serves as a gateway for the UE to connect to the core 5G network. The UE must connect to a gNB tower to gain access to the service offered by the 5G network.

c. **International Mobile Subscriber Identity (IMSI)** refers to a key which uniquely identifies a UE when it connects to a 5G network. This information is widely considered sensitive due to its ability to be traced to a specific user device.

### B. Brief overview on how a UE registers to a 5G network

As mentioned in [1], a UE will conduct a cell search within its vicinity to find any available gNB, - commonly referred to as cell towers or base station. More often, the UE will choose a base station based on its strength of its connection. When the UE has selected a cell tower, it will begin synchronizing with the base station. The UE will first conduct a downlink synchronization to determine the location of the base station by acquiring the Primary Synchronization Signals (PSS) and Secondary Synchronization Signals (SSS) and processing them. After this, the UE will send a preamble message to the base station, which responds to the UE with information allowing the UE to adjust it's timing with the base station. This synchronization process ensures that communications between the UE and base station are correctly synchronized and are not lost during transmission.

Once the UE has successfully synchronized, the UE will transmit a "Registration Request" to the base station indicating that it would like to connect to the 5G network for the first time.

Fig. 1. UE sending a Registration Request to the gNB

The UE will add several information in the request, some of which includes the UE's security capabilities, network capabilities, registration type (initial registration or reconnection) and its identifying information [1]. Some of the identifying information that can potentially be sent includes the device's International Mobile Subscriber Identity (IMSI).

Fig. 2. Content of a Registration Request sent by the UE to the gNB.

For the UE to successfully register with the 5G core network, it will have to authenticate itself, leading to the UE receiving an "Authentication Request".

Fig. 3. UE receiving an "Authentication Request" message.

To achieve this, the gNB will pass information obtained from the Registration Request to the Access and Mobility Management Function (AMF). The AMF will furthermore delegate this information over to the Authentication Server Function (AUSF), which works with the Unified Data Management to determine if this UE is authorized to join this network [1]. The UDF will then provide authentication vectors to the UE, which will include a set of randomly generated numbers to be used by the UE's authentication functions, and a list of expected responses. If the UE cannot successfully replicate the expected responses in its calculations, the UE will not be authorized to use the core network. All this information will be contained in the Authentication message sent to the UE.

Fig. 4.  UE receiving a "Security Mode Command" from the gNB

After authentication is successful, the gNB will send a "Security Mode Command" to the UE telling it which cipher and integrity protections will be used during all communications to ensure security.

The UE will acknowledge that it has accepted this command. At this point, the UE has successfully registered itself with the 5G network. There are a few more steps left for the UE to fully connect and begin transmitting on the network, but for the purposes of our discussion, we will ignore the last few steps of the process due to the lack of relevance.

III. IMPLICATIONS OF LOCATION TRACKING VULNERABILITY

As we have discussed in the introduction, malicious actors can utilize the unencrypted IMSI number transmitted during the Initial Registration of a UE to a base station to track a device on a network. The biggest implication of this setup is that end users will become easily susceptible to surveillance by adversaries including, but not limited to, overreaching government, law enforcement, criminals etc. These entities tend to use devices known as *IMSI catchers* or *Stingrays* to capture and monitor the IMSI numbers of UE devices and their locations.

A. Brief Backgound on how IMSI Catchers work

These devices generally work by impersonating a base station and sending strong signals to UE devices in the hopes that they will connect with the catcher. Given that UEs are designed to optimize their connectivity by connecting to base stations with the strongest signals [4], they will be tricked into connecting with the IMSI catcher. Like how a UE will react when connecting with a real cellular network, the UE will transmit it's IMSI in plaintext during the Initial Registration Request with the rouge base station [4]. Having obtained the IMSI, adversaries can use it to monitor the location of the UE through two primary ways:

- **Trilateration:** This is when an attacker sends a "RRC Connection Reconfiguration" to the target device. The command will contain the cell IDs of at least 3 neighboring cell towers as a parameter [26]. The attacker can enter the cell IDs of other rouge base stations/IMSI catchers. Normally this feature is used to adjust a configuration of the connection to the UE. However, the attacker is only interested in one field of the response. This field contains the frequency signal strength of the previously entered cell towers – which in our case are rouge base stations - [26], allowing the attacker to judge the location of the UE with high level of accuracies.

- **Using the LocationInfo-r10 feature:** Modern phones have the locationInfo-r10 as a field in the response of an "RRC Connection Reconfiguration". This filed contains the GPS coordinates of the UE, which will allow the attackers to pinpoint the location of the device with extreme accuracy.

There are more sophisticated IMSI catchers which can conduct man-in-the-middle attacks to actively intercept calls and SMS from target UE devices, but we will not explore this in detail as it is outside the scope of this paper.

B. Examples of IMSI catching in the real world

1. According to [4], FBI agents have testified in court that they have used IMSI catchers over 300 times in a decade. [4] also shows that 21 states are confirmed to have had their police stations knowingly used IMSI catchers in their operations and investigations. Police have also been noted to have raided homes which were incorrectly pinpointed as location of target UE devices [4].

2. In December 2014, protests broke out in Chicago after a grand jury decided not to indict Eric Garner, a black man who died in the hands of an NYPD police officer through the use of a choke hold. Police who were at the protests were spotted using a bizarre looking van to follow protesters, who reported that they were having trouble getting cell service. The hacktivist group, Anonymous, later released transcripts which showed that the police were using a IMSI catcher to monitor protesters. [4]

3. In 2014, it was reported by the Wall Street Journal that US Marshalls had Cessnas fitted with Stingrays to gather IMSI numbers and the locations of tens of thousands of mobile devices across a large area of land [4]. The Justice Department has refused to confirm or deny this information.

4. [27] reports that nearly 54 IMSI catchers were found in the United States, a sizeable number of which, were found belonging to foreign governments, suggesting that foreign governments may be tracking and monitoring the location of US residents.

These examples show that the easy retrieval of the plaintext IMSI have given adversaries an easy way to locate UE devices belonging to innocent end user, or end users who were not changed with any crime. The stingrays are dependent on IMSIs' being sent in plaintext to properly track them. This makes it even more critical that next generation networks come up with a mechanism to protect the IMSI number from interception to prevent location tracking by malicious actors.

## IV. HOW 5G NETWORKS MITIGATE THE LOCATION TRACKING VULNERABILITY

To combat the issue of unencrypted IMSI being transmitted from the UE to base station during initial registration, the 3rd Generation Partnership Program (3GPP), who design and standardize protocols for telecommunications, adopted TS 33.501, which adopted the Subscription Concealed Identifier (SUCI) as a key feature of modern 5G networks. Figure 6 shows a simplified version on how it works. In this new model, the IMSI of the UE is never sent in plaintext at any time during the initial registration with the network. As shown in Fig. 5, the SUCI approach stipulates that the UE will encrypt the MSIN of the IMSI, which is the most critical aspect of the IMSI, while leaving the MCC and the MNC in plaintext to allow for easier routing through the 5G network (From this point on, the term "encrypted IMSI" will be used to only indicate the ciphered MSIN).

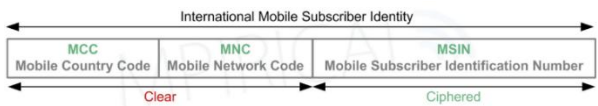

Fig. 5. The Composition of an IMSI

Source: Adapted from [5]

### A. Detailed discussion of the SUCI approach

In this new approach, the UE will be assigned a home network public key, which will be placed in the device's SIM card [8]. The UE will use the home network public key to generate a Diffie-Hellman shared key $S$. The UE will then use $S$ to encrypt the MSIN portion of the IMSI number using the principles of Elliptic Curve Integrated Encryption Scheme (ECIES), a modern-day encryption scheme which utilizes the mathematical concepts of elliptic curves to achieve strong encryption – we will discuss the finer details of this encryption scheme in a little bit-. There are two common ECIES encryption schemes used: Curve25519 and Secp256r1 and it is common to find these two encryption schemes built into popular 5G open-source implementations, including Open5GS. The network operators can choose which of the encryption schemes they want to use, but both choices are resoundingly secure.

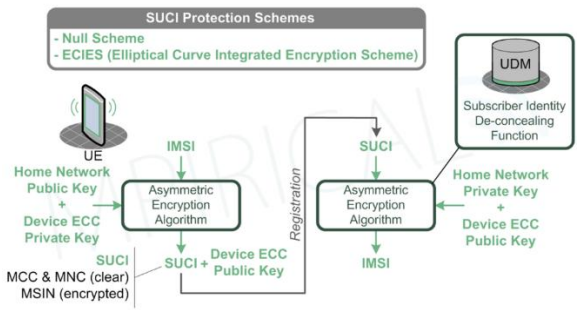

Fig. 6. The SUCI Approach in 5G Networks

Source: Adapted from [25]

Once the MSIN portion of the IMSI has been fully encrypted, the UE will start creating the Initial Registration Request message to send to the gNb, like how it does in previous generations. In the message, the UE will specify the following:

a. **Format:** Using an integer, the UE will indicate what type of long-term identifier it is using. In most cases, a 0 indicates an IMSI, while other numbers may be mapped to identifier such as GUTI, TMSI etc.

b. **Public Land Mobile Network:** The Public Land Mobile Network will store the MCC and MNC, which both identify the country of origin and the mobile operator of the UE. This information is store unencrypted.

c. **Protection Scheme ID:** This field will contain an integer indicating which encryption scheme the UE is using to encrypt the IMSI. A 0 will indicate a 'Null protection scheme" while a 1 or a 2 will indicate a specific ECIES encryption algorithm.

d. **Output:** This field will contain the information needed by the gNb to register the UE on the 5G network. It contains the following:

   a. **Home Network Public Key:** This is the public key that was provisioned to the UE by the network.
   
   b. **Encrypted MSIN:** This is where the UE will send the fully encrypted MSIN to the gNb.
   
   c. **MAC Number:** This is a Message Authorization Code that it sent.

Once the UE sends the message to the core network, the Unified Data Management (UDM) will be tasked with unencrypting the IMSI sent. In 5G networks, the UDM has a function named Subscriber Identity Deconcealing Function [25] which will decrypt the IMSI. As mentioned in Fig. 6, it will use the home networks private key, which is provisioned to the network, and the UE's public key, which is contained in the message received by the base station.

### B. Elliptic Curve Integrated Encryption Scheme

Given the major role of the elliptic curve encryption scheme in the protection of the IMSI, it makes sense to briefly go over how elliptic curve encryption and the characteristics that make it ideal for securing our IMSI.

It is important to note that the Elliptic Curve Integrated Encryption Scheme is considered a hybrid encryption scheme since it utilizes multiple facets of modern-day encryption to achieve robust security [31]. Some of these facts include:

- **Public Key Cryptography:** The scheme utilizes a public/private key setup where all participants in the communication have public keys and private keys to facilitate the secure communications.

- **Key Exchange:** The scheme uses a Diffie Hellman key exchange technique to generate a shared key which will be used for both encryption and decryption by the sender and the receiver.
- **Symmetric Encryption:** The scheme will use a symmetric encryption algorithm to secure the plaintext being sent. A common one used is AES.
- **MAC Verification:** The scheme uses a verification system to ensure that the data has not been changed in transit. We will not investigate this in detail.

Many encryption schemes also offer these facets, but what makes ECIES more secure is the mathematics behind Elliptic Curve Cryptography (ECC).

According to [11], elliptic curves are curves defined by an equation. The equations take the standard form.

$$y = x + ax + b \qquad (1)$$

where $a, b$ are some constants. We can define $F$ as the Galois field of mod-$p$ remainders where $p$ is a prime number and $p > 3$ [11]. $F$ will only contain integer coordinates over the square matrix of size p x p, an example of which is shown in Fig.7. In $F$, we can use Elliptic Curve Multiplication to find other points in the field which are on our defined elliptic curve.

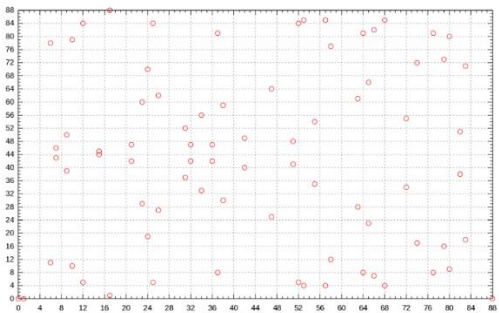

Fig. 7. The set of points of an example elliptic curve $y^2 = x^3 - x$ over a finite field $F_{89}$. When applied to the field, the points on the curve are separated and scattered due to the mod operation, mixed with points not on the curve.

Source: Adapted from [30]

Elliptic Curve Cryptography uses these basic principles to achieve a level of security. If we have a fixed point on the elliptic curve $G$ and multiply it by an integer $k$ (which can be considered as a private key), we can obtain a point $P$, which is guaranteed to be on the elliptic curve (this can be considered a corresponding public key) [31]. This computation is considered to be extremely easy and the product can be found with minimal computation. However, the same cannot be said about the reverse. It is extremely difficult to find $k$ given $P$ and $G$. Infact, if $k$ is large enough, it is considered computationally infeasible. This gives rise to the **Elliptic Curve Discrete Logarithm Problem**, which is defined as the following [31]:

- By given elliptic curve over a finite field $F_p$ and generator point $G$ on the curve and point $P$ on the curve, find the integer $k$ such that $P = k * G$

As long as the elliptic curve and the finite field $F_p$ are chosen wisely, this problem has no feasible solution, which is what the encryption scheme depends on in order to provide robust security.

V. SIMULATION OF THE SUCI APPROACH

*A. Simulation Design and Assumptions*

The purpose of this simulation is to demonstrate on a small scale how the IMSI is encrypted in a UE and decrypted on the 5G network side. The simulation was written in Python and uses two open source Python libraries namely CryptoMobile and PyCrate. The simulation assumes that home network does not have a public and private key provisioned and therefore manually creates a pair for it. The simulation also assumes that we are using an IMSI with a MCC of 242, MNC of 01, and an MSIN of 534567890. The full IMSI will be 24201-534567890 which corresponds with a random Norwegian Telecom operator. Finally, the simulation will assume that there is a mechanism in place to transfer data between the UE and the home network. The simulation does not implement any infrastructure that facilitates such data transfer.

*B. Simulation*

- **UE Side:** The UE will first create a message object to send for the Initial Registration Request with the 5G network. It will fill the message with the necessary parameters, including – for now - the unencrypted MSIN portion of the IMSI number. It will set the encryption scheme as null indicating no encryption for now.

Fig. 8. The current output of the message object created by the UE. The MSIN portion of the IMSI has not been encrypted yet.

The UE will then select it's elliptical curve encryption scheme of choice. The UE will set the "ProtSchemeID" field as 1, indicating that it will encrypt the MSIN portion using the curve25519 encryption scheme. The UE generates a shared key using the home network's public key and uses that to fully encrypt the MSIN portion using the CryptoMobile library. The Python implementation used in the simulation utilizes AES in the backend to encrypt the MSIN, with the shared key acting as the symmetrical key.

Once the encryption has been done, the UE will create new fields in the "Output" section where it will then append,

the it's public key, MSIN, and MAC number respectively with all of them encrypted.

```
### 5GSIDSUPI ###
  <spare : 0>
  <Fmt : 0 (IMSI)>
  <spare : 0>
  <Type : <FGSIDTYPE.SUPI: 1> (SUCI)
  ### Value : 0 -> SUCI_IMSI ###
    <PLMN : 24201 (Norway.Telenor)>
    <RoutingInd : 0000>
    <spare : 0x0>
    <ProtSchemeID : 1 (ECIES scheme profile A)>
    <HNPKID : 0>
    ### Output : 1 -> _SUCI_ECIESProfA ###
      <ECCEphemPK : 0x1942e5a43c2bd42d88a52f4039ce08b40ec00f808c7496cf2ffefc35e081ee6b>
      <CipherText : 0x99af8f77fe>
      <MAC : 0xac2fadc8b7d9fe06>
```

Fig. 9. The current output of the message object on the UE. The MSIN portion of the IMSI has been encrypted using curve25519 encryption.

- **Home Network Side:** The home network will receive the identifying information as part of the Initial Registration Request. As shown in Fig. 8, the MCC and the MNC are in the clear while the MSIN is encrypted. To unencrypt it, the home network will have to generate the shared key using the UE's public key. It will then decrypt the MSIN and the MAC number using functions provided in the CryptoMobile library.

```
'24201-534567890'
```

Fig. 10. The fully decrypted IMSI, showing the already unencrypted MCC and MNC with the newly unencrypted MSIN. This matches with the IMSI we started off with

## VI. DISCUSSION AND LIMITATIONS

The SUCI model has proven to be effective in masking the MSIN portion of the IMSI, allowing the protection of the UE. It guarantees that during the Initial Registration of a UE to the 5G network, the UE will not transmit the IMSI in plaintext. The model fulfils the three criteria set out in the introduction of the paper namely:
- The SUCI model ensures that the MCC and the MNC are unencrypted to ensure the ease of data routing throughout the network, while ensuring the MSIN is securely encrypted. This is achieved by splitting the MSIN from the MCC and the MNC and reconstructing the IMSI on the network side after decryption. This hampers the abilities of IMSI catchers since any interception of the IMSI in transit would be encrypted and without the ability to compute the shared key, the IMSI catcher would not be able to unencrypt it.
- The SUCI model securely encrypts the MSIN portion of the IMSI using the ECIES encryption scheme, which combines Elliptic Curve Diffie Hellman key exchange and symmetric encryption (using AES) to achieve strong encryption. If the curve and the finite field is chosen wisely, it would be practically impossible to compute the shared key using brute force.
- The SUCI model uses a public key infrastructure to conduct the Diffie Hellman key exchange. The 5G network is provisioned a public and private key by the network operator and the UE is also provisioned a public and private key. The UDM also now has a Subscriber Identity Deconcealing Function to unencrypt the MSIN portion of the IMSI number and reveal the identity of the UE.

Despite this, this security standard in 5G networks has fallen under criticism for perceived holes in its security model. [12] argues that the model gives the network operators too much discretion in enabling the encryption scheme protections. Operators have the option to disable any protection on the IMSI during transit, as seen during the simulation where one could set "ProtSchemeID" to 0, indicating Null Protection Scheme. This is by design in case of the need for emergency access [12], but opens the door for complacency and could lead to network operators disabling protections on the IMSI for any number of reasons. Operators also have the option to not routinely provision the home network public key on the UE's SIM. In fact, they can decide to not provision it at all [12]. If the UE cannot find the home network's public key in the SIM, the null scheme protection system will be used by default [12], defeating the purpose of the renewed security standards in 5G. This discretionary model represents a single point of failure in the 5G security model where one mistake by a network engineer could put the privacy of thousands of end user at risk.

This paper recommends that 3GPP should disable null protection schemes and enforce ECIES encryption as the default. Admittedly, this could disrupt the ability for UEs to conduct calls during emergency situations and therefore more research must be done to find an alternative way to place emergency calls without exposing the IMSI of the UE.

After discussing the core functionalities of the SUCI model in securing the IMSI, it is essential to consider its applications and implications in the broader context of modern communication technologies based on machine learning [32-40]. The advancement of machine learning presents a unique opportunity to enhance network security in 5G. By employing ML algorithms, we can create more dynamic and responsive security measures, particularly in the IMSI transmission process, making it more resilient against sophisticated cyber threats.

Furthermore, the intersection of 5G with other prevalent communication technologies [41-49] like WiFi, Bluetooth Low Energy (BLE), and Zigbee adds another layer of complexity to the security landscape. The amalgamation of these technologies within the 5G framework necessitates a security protocol that is not only robust for the cellular network but also adaptable to the unique requirements of these varied communication standards. This adaptability is critical in ensuring the secure and seamless functioning of interconnected systems, which are increasingly prevalent in scenarios ranging from smart home applications to complex industrial IoT setups [50-64].

In this evolving communication ecosystem, the role of security [65-70], particularly the SUCI model, becomes even more paramount. It must provide a secure foundation that supports the diverse and concurrent communication needs of these technologies while maintaining the integrity and confidentiality of sensitive data like the IMSI.

## VII. CONCLUSION

This paper discussed the issue surrounding location tracking vulnerability and how transmitting an unencrypted IMSI by the UE during initial registration with the network can lead to possible end user location tracking by malicious actors. The paper discusses how fake base stations such as IMSI catchers can exploit this design feature in previous generation networks to facilitate location tracking, showcasing the need for a new approach for 5G networks to identify UE devices. The paper explored the SUCI model, which was introduced in specification TS 33.501 as a step in the right direction in combatting this vulnerability. The paper explored how the SUCI model uses Elliptic Curve Integrated Encryption Scheme (ECIES) to securely generate a shared key which is used by the UE to encrypt the MSIN portion of IMSI, leaving the MCC and MNC in plaintext, before being decrypted by the core 5G network upon receipt. Despite the excellent protection offered by this approach, the paper notes that deficiencies are still present and offers a few recommendations to help improve the model even further.